\documentclass[twocolumn,showpacs,superscriptaddress,twoside,prl]{revtex4}
\usepackage{amsmath,amssymb,url}
\usepackage{graphicx,epstopdf,bm}
\usepackage{epsfig}
\usepackage{hyperref}
\usepackage{amsmath}
\usepackage{amssymb}
\usepackage{makeidx}
\usepackage{multirow}
\usepackage{xcolor}
\usepackage{color}
\usepackage[caption=false]{subfig}

\newcommand{\blue}{\color[rgb]{0,0,0.8}}

\begin{document}
\title{Double-double electromagnetically induced transparency with amplification}
\author{Hessa M. M. Alotaibi}
\email{hmalotai@ucalgary.ca}
\affiliation{Institute for Quantum Science and Technology, University of Calgary, Alberta, Canada T2N 1N4}
\affiliation{Public Authority for Applied Education and Training, P.O. Box 23167, Safat 13092, Kuwait }
\author{Barry C. Sanders}
\affiliation{Institute for Quantum Science and Technology, University of Calgary, Alberta, Canada T2N 1N4}
\affiliation{Hefei National Laboratory for Physical Sciences at Microscale,
        University of Science and Technology of China, Anhui 230026, China}

\begin{abstract}
We show that an alkali atom with a tripod electronic structure
can yield rich electromagnetically induced transparency phenomena even at room temperature.
In particular we introduce double-double electromagnetically induced transparency wherein
signal and probe fields each have two transparency windows.
Their group velocities can be matched in either the first or second pair of transparency windows.
Moreover signal and probe fields can each 
experience coherent gain in the second transparency windows.
We explain using a semi-classical-dressed-picture to
connect the tripod electronic structure to a double-$\Lambda$ scheme.
\end{abstract}

\date{\today}
\maketitle

Electromagnetically induced transparency (EIT) exploits interfering electronic transitions in a medium to eliminate absorption and dramatically modify dispersion over a narrow frequency band
with applications including slow light, reduced self-focusing and defocusing~\cite{Harris1997},
and quantum memory~\cite{Lvovsky2009}.
Microscopically, a three-level~$\Lambda$ electronic structure suffices to explain EIT.
Double EIT (DEIT) extends EIT to creating two simultaneous transparency windows,
one for a ``signal'' and the other for a ``probe'' field, 
with the aid of a third ``coupling'' field~\cite{ Rebic2004, Joshi2005, Wang2006, Ottaviani2006, MCL08}.
DEIT is valuable for coherent control and enabling long-lived nonlinear interactions between weak fields,
which could enable deterministic all-optical two-qubit gates for quantum computing.

\begin{figure}
\centering
    \subfloat[\label{subfig-barestate}]{%
      \includegraphics[width=0.47\columnwidth]{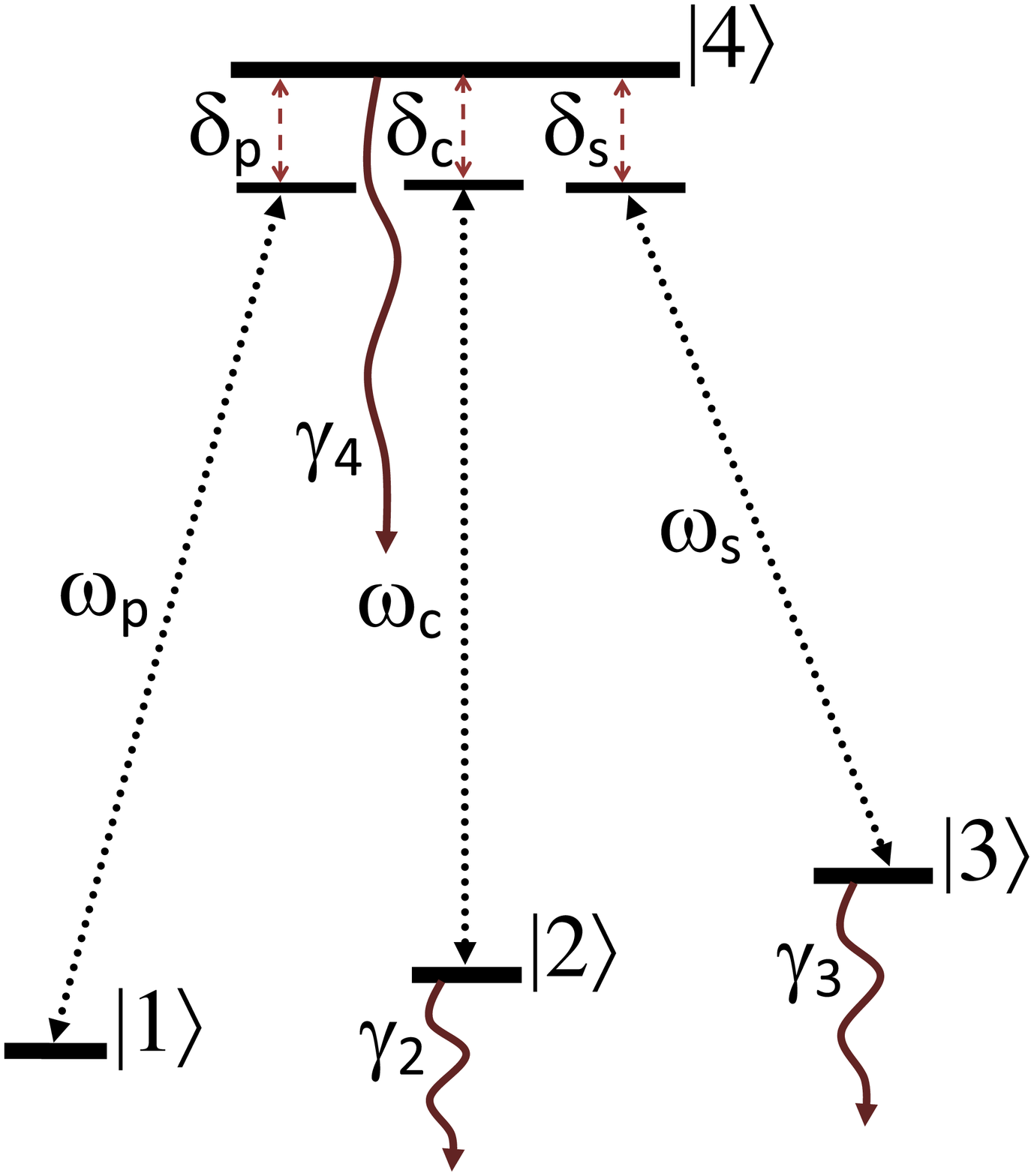}
    }
   \hfill
    \subfloat[\label{subfig-dispersiony}]{%
      \includegraphics[width=0.47\columnwidth]{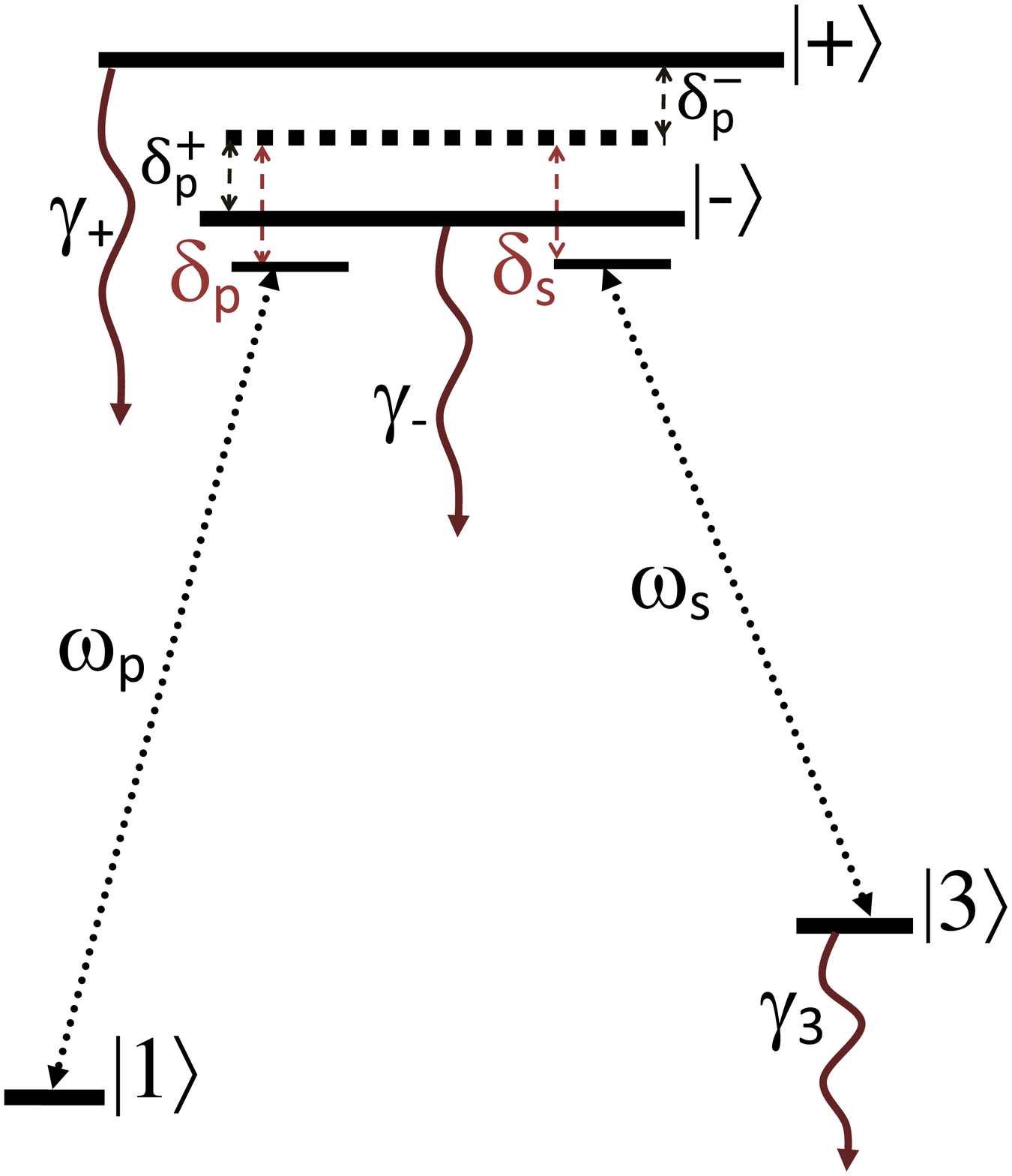}
    }
    \caption{(a)~Four-level tripod electronic structure with high-energy state $|4\rangle$
		and lower-energy levels $|1\rangle$, $|2\rangle$, and $|3\rangle$ in order of 
		increasing energy. Transitions are driven by probe (p), coupling (c) and 
		signal~(s) fields with frequencies $\omega_\text{x}$ and detunings $\delta_\text{x}$
		with x$\in\{\text{p,c,s}\}$. 
		Decay rates for level~$|i\rangle$ are $\gamma_i$ for $i\in\{2,3,4\}$.
		(b)~Same atom in semi-classical-dressed-picture for strong c-field,
		which corresponds to a double-$\Lambda$ level structure.
		Levels~$|2\rangle$ and~$|4\rangle$ are hybridized into $|\pm\rangle$.}
 \label{fig:tripod}
  \end{figure}

Whereas the $\Lambda$ scheme suffices to explain EIT,
DEIT requires at least four levels.
The tripod ($\pitchfork$) scheme~\cite{Rebic2004}, which has one upper
and three lower levels as shown in Fig~\ref{fig:tripod}(a),
is one such four-level scheme.
This scheme can be reframed in the semi-classical-dressed-picture~\cite{Ber96, Fleischhauer2005} shown in Fig.~\ref{fig:tripod}(b),
which has two lower ($|1\rangle$ and~$|3\rangle$) and two upper ($|\pm\rangle$) levels
after eliminating the strong coupling~(c) field.

This semi-classical-dressed model of the $\pitchfork$ scheme corresponds effectively to a 
double~$\Lambda$ system,
and double~$\Lambda$ schemes have been studied experimentally~\cite{VAFL07}.
With our semi-classical-dressed analogy,
we show that this $\pitchfork$ electronic structure exhibits rich hitherto-unnoticed EIT phenomena,
namely what we now call double DEIT (DDEIT).
Our DDEIT phenomenon has the property that both the signal and the probe fields can each have two EIT windows
given the right parameter choices.

One particular aspect of our system, namely the second EIT window for the probe, has been predicted~\cite{Paspalakis2002}
and observed experimentally~\cite{Li2007,MCL08}, but this previously observed effect
corresponds only to one aspect of our system,
namely a double window for the probe and not to our full DDEIT for both signal and probe fields.
Moreover, these new second EIT windows for each of the signal and probe fields exhibit coherent gain,
which has not previously been expected.

We now reprise the dynamics of the driven~$\pitchfork$ atom~\cite{Rebic2004}.
For $\hbar\equiv 1$ and $\hat{\sigma}_{\imath\jmath}:=|\imath\rangle\langle\jmath|$,
the free Hamiltonian is $\hat{H}_0=\sum_{\imath=1}^4\omega_\imath\sigma_{\imath\imath}$.
For $\omega_{\imath\jmath}:=\omega_\imath-\omega_\jmath$,
the $\pitchfork$ atom is driven by a probe field with frequency~$\omega_\text{p}=\omega_{41}-\delta_\text{p}$,
a coupling field with frequency~$\omega_\text{c}=\omega_{42}-\delta_\text{c}$,
and a signal field with frequency~$\omega_\text{s}=\omega_{43}-\delta_\text{s}$.
In terms of Rabi frequencies $\Omega_\text{x}$ for $x\in\{\text{p},\text{c},\text{s}\}$,
the driving Hamiltonian is
$	\hat{H}_\text{dr}(t)
		=\frac{1}{2}\left(\Omega_\text{p}\text{e}^{i\omega_\text{p}t}\hat{\sigma}_{14}
			+\Omega_\text{c}\text{e}^{i\omega_\text{c}t}\hat{\sigma}_{24}
			+\Omega_\text{s}\text{e}^{i\omega_\text{s}t}\hat{\sigma}_{34}
			+\text{hc}\right)$
for hc the Hermitian conjugate.

Under a rotating-frame transformation with respect to
$\hat{A}=3\delta_\text{p}\hat{\sigma}_{11}
	+(2\delta_\text{p}+\delta_\text{c})\sigma_{22}
	+(2\delta_\text{p}+\delta_\text{s})\hat{\sigma}_{33}
	+2\delta_\text{p}\hat{\sigma}_{44}$,
the resultant time-independent Hamiltonian is~\cite{Rebic2004}
$\hat{H}=\delta_\text{pc}\sigma_{22}+\delta_\text{ps}\sigma_{33}+\delta_\text{p}\sigma_{44}
	+\left(\Omega_p\sigma_{41}+\Omega_c\sigma_{42}
	+\Omega_s\sigma_{43}+\text{hc}\right)/2$
with~$\delta_\text{xy}:=\delta_\text{x}-\delta_\text{y}$.
The Lindblad master equation is
\begin{align}
	\dot{\rho}
		=&i[\rho,\hat{H}]+\sum_{\imath<\jmath}^4{\frac{\gamma_{\jmath\imath}}
			{2}(\sigma_{ij}\rho\sigma_{ji}-\sigma_{jj}\rho-\rho\sigma_{jj}})\nonumber\\
	&+\sum_{j=2}^4{\frac{\gamma_{\phi j}}{2}(\sigma_{jj}\rho\sigma_{jj}-\sigma_{jj}\rho-\rho\sigma_{jj})}
\label{eq:masterequation}
\end{align}
including spontaneous emission and dephasing.
The decay rates depicted in Fig.~\ref{fig:tripod}(a) are
$\gamma_j:=\sum_{i<j}(\gamma_{ji}+\gamma_{\phi j})$,
and the steady-state density-matrix ($\bar{\rho}$) solution is known~\cite{Rebic2004}.

\begin{figure}
	\includegraphics[width=.92\columnwidth]{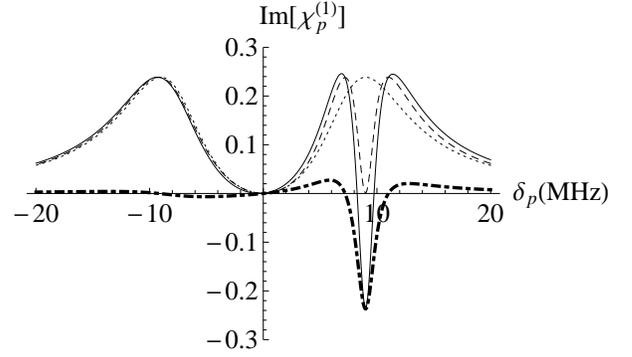}
	\caption{
		Absorption~Im$[\chi_{p}]$ 
		vs probe detuning $\delta_\text{p}$ 
		for $\gamma_4=18$MHz, $\gamma_3=10$kHz, 
		$\gamma_2=40$kHz, $\Omega_c=\gamma_4$, 
		$\Omega_s=0.3\gamma_4$, $\delta_s=0.5\Omega_c$, and $\delta_c=0$
		with all terms included (solid),
		with $\rho_{43}^{(0)}\equiv 0$ imposed (dash),
		the difference between these two cases (dot-dash),
		and the case that $\Omega_\text{s}\equiv 0$ (dot).
		}
\label{fig:probeplots}
\end{figure}

Given~$\bar{\rho}$, optical susceptibility of the $\pitchfork$ medium can be calculated.
Here we are concerned only with linear optical susceptibility~$\chi^{(1)}$
so we calculate ~$\bar{\rho}_{14}$ to lowest order in $\Omega_\text{p}$
but retain $\Omega_\text{c,s}$ to all orders
and similarly calculate ~$\bar{\rho}_{34}$ to lowest order in $\Omega_s$
while retaining $\Omega_\text{c,p}$ to all orders.
nonlinear susceptibility is a topic for future study.
We first consider the probe-field case, and the signal case is similar.
The part of $\bar{\rho}_{14}$ that is linear in~$\Omega_\text{p}$ is
\begin{equation}
	\bar{\rho}^\text{lin}_{14}
		=\frac{\text{i}\left(\rho_{11}-\rho_{44}\right)
			+\frac{\Omega_\text{s}}{\gamma_3-2i\delta_\text{ps}}\rho^{(0)}_{43}}
			{\gamma_4-2\text{i}\delta_\text{p}-\frac{\Omega_c^2}{\gamma_2-2i\delta_\text{pc}}
			+\frac{\Omega_\text{s}^2}{\gamma_3-2\text{i}\delta_\text{ps}}}
			\Omega_\text{p}
\label{eq:rho14}
\end{equation}
with
\begin{equation}
	\bar{\rho}^{(0)}_{43}
		:=\left.\rho_{43}\right|_{\Omega_\text{p}=0}
		=\frac{-\text{i}\Omega^*_s (\rho_{33}-\rho_{44})}
		{\gamma_3+\gamma_4+2\text{i}\delta_\text{s}
			+\frac{\Omega_c^2}{\gamma_3+\gamma_2+2\text{i}\delta_\text{sc}}}.
\label{eq:rho34}
\end{equation}
For an atomic gas in three dimensions with~$\mathcal{N}$ the atomic density
and~$\bm{d}_{14}$ the dipole moment,
the linear optical susceptibility is~\cite{Rebic2004}
\begin{equation}
	\chi^{(1)}_\text{p}
		=\frac{\mathcal{N}\left|\bm{d}_{14}\right|^2}{\epsilon_0}
			\frac{\bar{\rho}^\text{lin}_{14}}{\Omega_\text{p}}.
\end{equation}
Our Eq.~(\ref{eq:rho14}) generalizes the previously known expression,
as the earlier equation ignores the signal field effect
on the probe linear optical response ($\Omega_\text{s}\equiv 0$)
because the focus was on the equal-detuning special case $\delta_p=\delta_s=\delta_c=0$~\cite{Rebic2004}.
The validity of ignoring $\Omega_\text{s}$ is evident in Fig.~\ref{fig:probeplots}
near $\delta_\text{p}=0$ but definitely not away from that region where the signal plays a key role in interference
and is needed for DDEIT.

We calculate and plot Im$[\chi^{(1)}_p]$ (absorption) in Fig.~\ref{fig:probeplots}.
In order to explain the parameter choices in Fig.~\ref{fig:probeplots},
we refer to Fig.~\ref{fig:tripod}(a).
Specifically we consider~$^{87}$Rb and assign~$|1\rangle$ to the $5S_{1/2}$ level
with $F=1$ and $m_F=0$,
$|2\rangle$ to the $5S_{1/2}$ level with $F=2$ and $m_F=-2$
and~$|3\rangle$ to the~$5S_{1/2}$ level with~$F=2$ and $m_F=0$.
Level~$|4\rangle$ corresponds to the~$5P_{1/2}$ level with $F=2$ and $m_F=-1$.

For this atom $\omega_{4\imath}=2.369\times 10^{15}$Hz with  right-circular polarization
for $\imath=2$ and left-circular polarization for $\imath=1, 3$.
The decay rates~\cite{MCL08} and field strengths are given in the caption of Fig.~\ref{fig:probeplots}.
The atomic density is $10^{14}\text{cm}^{-3}$.

Figure~\ref{fig:probeplots} exemplifies the features inherent in Eq.~(\ref{eq:rho14}).
First consider the case that $\Omega_\text{s}\equiv 0$,
which decouples~$|3\rangle$
from the dynamics and restores ordinary $\Lambda$-atom EIT.
The  semi-classical-dressed picture of Fig.~\ref{fig:tripod}(b) clarifies the dynamics
where we introduce two dressed states $|\pm\rangle$.
The $\Omega_\text{s}\equiv 0$ line in Fig.~\ref{fig:probeplots} shows
two EIT absorption peaks at $\delta_\text{p}^\pm$
corresponding to $|1\rangle\leftrightarrow|\pm\rangle$ transitions, respectively.

\begin{figure}
	\includegraphics[width=.9\columnwidth]{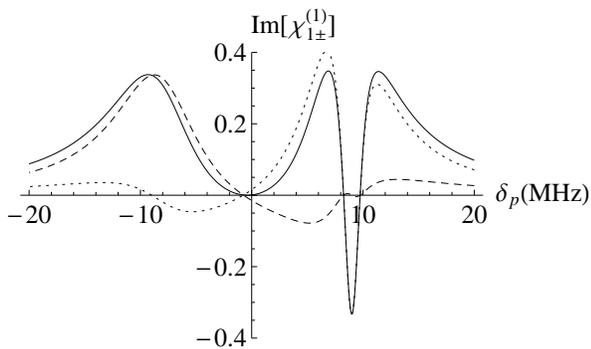}
	\caption{
		Im$(\chi_{1\pm})$ 
		vs probe detuning $\delta_\text{p}$ 
		for $\gamma_4=18$MHz, $\gamma_3=10$kHz, 
		$\gamma_2=40$kHz, $\Omega_c=\gamma_4$, 
		$\Omega_s=0.3\gamma_4$, $\delta_s=0.5\Omega_c$, and $\delta_c=0$
		with $\bar{\rho}_{1+}^{\text{lin}}$ (dash), $\bar{\rho}_{1-}^{\text{lin}}$ (dot)
		and $\bar{\rho}_{1+}^{\text{lin}}+\bar{\rho}_{1-}^{\text{lin}}$ (solid).
		}
\label{fig:dressedplots}
\end{figure}

Mathematically the  semi-classical-dressed picture is obtained by the unitary transformation~\cite{Ber96, Fleischhauer2005}
\begin{equation}
	\rho\mapsto U\rho U^\dagger,\;
	U=\begin{pmatrix}
		1&0&0&0 \\
		0&\vartheta &0&\vartheta\varsigma\\
		0&0&1&0\\
		0& -\vartheta\varsigma^*&0&\vartheta
		\end{pmatrix}
\end{equation}
with $\varsigma:=\frac{\sqrt{|\Omega_\text{c}|^2+\delta^2_\text{c}}+\delta_\text{c}}{\Omega_\text{c}}$
and $ \vartheta:=\frac{1}{\sqrt{1+|\varsigma|^2}}$.
In this  semi-classical-dressed basis
\begin{equation}
\label{eq:1-lin}
	\bar{\rho}^\text{lin}_{1-}=\left(\vartheta\varsigma+\frac{i\vartheta\Omega^*_\text{c}}
		{\gamma_2-2i\delta_\text{pc}}\right)\bar{\rho}^\text{lin}_{14},
\end{equation}
and
\begin{equation}
\label{eq:1+lin}
	\bar{\rho}^\text{lin}_{1+}=\left(\vartheta-\frac{i\vartheta\varsigma^\ast \Omega^*_\text{c}}
		{\gamma_2-2i\delta_\text{pc}}\right)\bar{\rho}^\text{lin}_{14},
\end{equation}
which are plotted in Figs.~\ref{fig:dressedplots}.
Equations~(\ref{eq:1-lin}) and~(\ref{eq:1+lin}) are useful because the undressed state
$\bar{\rho}_{14}^{\text{lin}}$ corresponds to interfering transitions to $\bar{\rho}_{1\pm}^{\text{lin}}$.

For $\Omega_\text{s}\neq 0$,
we see in Fig.~\ref{fig:probeplots} that the second absorption peak
at $\delta_\text{p}^+$ is split by a transparency window with negative absorption, i.e., gain.
This splitting of the second peak is due to the formation of a double-$\Lambda$ electronic structure~\cite{VAFL07} shown in Fig.~\ref{fig:tripod}(b).
Specifically, level $|+\rangle$ gives the absorption peak at $\delta_\text{p}^-$,
but the peak at $\delta_\text{p}^+$ is split by competing transitions
$|1\rangle\leftrightarrow|-\rangle$ and $|3\rangle\leftrightarrow|-\rangle$.

This explanation of competing transitions elucidates the splitting of the $\delta_\text{p}^+$ peak but not the presence of gain in the second EIT window ($\delta_p=\delta_s$).
In Fig.~\ref{fig:dressedplots} gain in $\bar{\rho}_{1+}^{\text{lin}}$
is evident over a wide
domain of~$\delta_\text{p}$  but cancels everywhere in the sum
$\bar{\rho}_{1+}^{\text{lin}}+\bar{\rho}_{1-}^{\text{lin}}$ except in the narrow second EIT window.
This gain is due to off-resonant driving to one of the upper levels.
Both Im$\bar{\rho}_{1\pm}^{\text{lin}}$ contribute to the probe susceptibility,
which is proportional to $\bar{\rho}_{14}^{\text{lin}}$.
The gain for the $|1\rangle\leftrightarrow|+\rangle$ is overwhelmed by the loss
due to driving the $|1\rangle\leftrightarrow|-\rangle$ transition on or near resonance.
This loss overwhelms the gain leading to no gain for the probe transition except in a narrow window as
seen in Fig.~\ref{fig:dressedplots}.

As this gain is initially surprising,
we investigate further using the undressed picture of Fig.~\ref{fig:tripod}(a).
In the undressed picture
the population in $|2\rangle$ and $|4\rangle$ vanish only when there is no decay from $|3\rangle$,
i.e., $\gamma_{3}=0$. 
As $\gamma_4>>\gamma_{3}$,
any  population pumped by the coupling field to~$|4\rangle$
will then decay to~$|1\rangle$  and~$|3\rangle$. 
Condition $\gamma_4>>\gamma_{3}$ ensures that,
at steady state, the  population of $|1\rangle$ or $|3\rangle$ 
exceeds the combined population of $|2\rangle$  and $|4\rangle$.
Thus, gain is not due to population inversion in $|4\rangle$ or due to hidden population inversion in $|\pm\rangle$
but rather due to quantum coherence inherent in $\rho_{43}^{(0)}$,
which is due to signal-field driving.

Mathematically, gain due to signal-driven coherence is evident in Eq.~(\ref{eq:rho14}),
which is a sum of two terms: one proportional to population difference $\rho_{11}-\rho_{44}$
and the other proportional $\rho_{43}^{(0)}$.
Gain occurs at $\delta_\text{p}=\delta_\text{s}$ for which the imaginary part of the first term is positive
and the imaginary part of the second term is negative.
Probe gain arises due to signal-driven coherence via $|1\rangle\leftrightarrow|3\rangle$ coherence:
$\dot{\rho}_{13}
	=(-\frac{1}{2}\gamma_3+i\delta_\text{ps})\rho_{13}-\frac{i}{2}(-\rho_{14}\Omega^*_\text{s}+\rho^{(0)}_{43}\Omega_\text{p})$,
which shows that this coherence is responsible for coupling the signal- and probe-driven transitions.
This $|1\rangle\leftrightarrow|3\rangle$ coherence is crucial to establish the requisite interfering channels
in order to enable gain to outweigh the effects of absorption~\cite{Koc92, CB96}. 
 
\begin{figure}
	\includegraphics[width=.9\columnwidth]{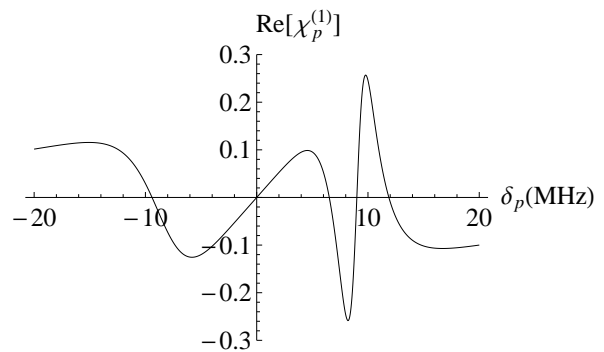}
	\caption{
		Re$[\chi_{p}]$
		vs probe detuning $\delta_\text{p}$ 
		for $\gamma_4=18$MHz, $\gamma_3=10$kHz, 
		$\gamma_2=40$kHz, $\Omega_c=\gamma_4$, 
		$\Omega_s=0.3\gamma_4$, $\delta_s=0.5\Omega_c$, and $\delta_c=0$.
		}
\label{fig:Rprobeplots}
\end{figure}

Probe dispersion is shown in Fig.~\ref{fig:Rprobeplots}.
Group velocity scales inversely with slope,
which is approximately constant in each of the two EIT windows.
For detuning~$\delta_\text{p}$ chosen at the center of each window,
dispersion is zero so the ratio of group velocities for each EIT window is thus the inverse of the 
ratio of the slopes for each window. 
From the plot, group velocity at the first window evidently exceeds group velocity at the second window
for the given parameters.

We now have a clear understanding of both DDEIT and gain for the probe field,
and now we investigate DDEIT and gain for the signal field.
The part of~$\bar{\rho}_{34}$ that is linear in $\Omega_\text{s}$ is
\begin{equation}
	\bar{\rho}^\text{lin}_{34}=\frac{\text{i}(\rho_{33}-\rho_{44})+\frac{\Omega_p}
	{\gamma_3+2\text{i}\delta_\text{ps}}\rho^{(0)}_{41}}
	{(\Gamma_{43}-2i\delta_\text{s}+\frac{\Omega_c^2}{\Gamma_{32}-2\text{i}\delta_\text{sc}}
	+\frac{\Omega_\text{p}^2}{\gamma_3+2\text{i}\delta_\text{ps}})}
	\Omega_\text{s}
\label{eq:Roh34}
\end{equation}
for
\begin{equation}
	\bar{\rho}^{(0)}_{41}
		=\left.\rho_{41}\right|_{\Omega_\text{s}=0}
		=-\frac{\text{i}\Omega^*_\text{p}(\rho_{11}-\rho_{44})}
			{\gamma_{4}+2\text{i}\delta_\text{p}+\frac{\Omega_c^2}{\gamma_{2}+2\text{i}\delta_\text{pc}}}.
\label{roh140Eq}
\end{equation}
\begin{figure}
\centering
    \subfloat[Absorption\label{subfig-absorption}]{%
      \includegraphics[width=0.48\columnwidth]{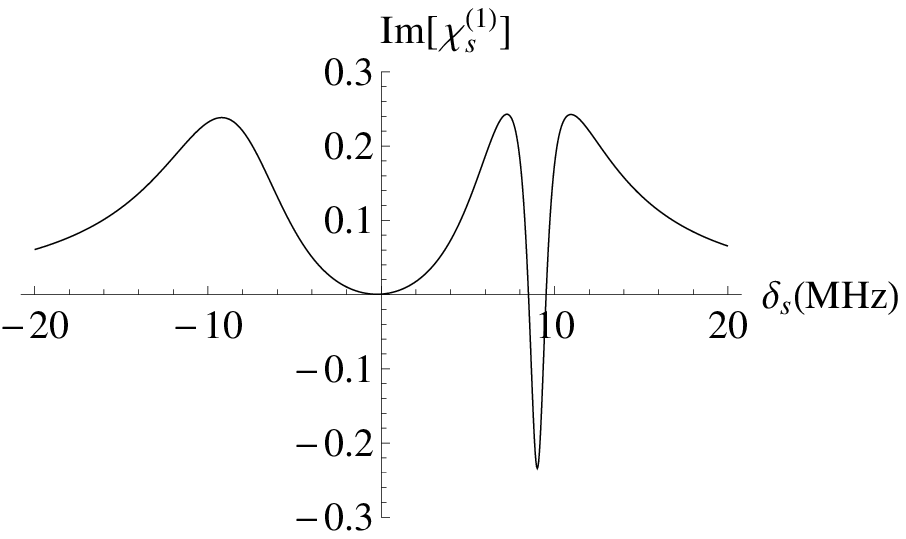}
    }
   \hfill
    \subfloat[Dispersion\label{subfig-dispersiony}]{%
      \includegraphics[width=0.48\columnwidth]{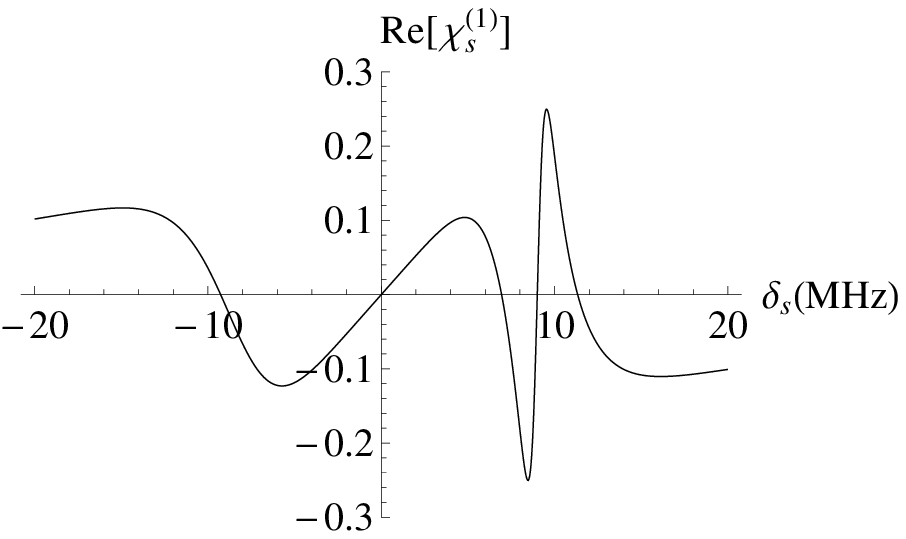}
    }
    \caption{
    	(a)~Absorption  and (b) Dispersion as a function of the probe detuning $\delta_\text{p}$ 
	 with $\gamma_3=10$kHz, $\gamma_4=18$MHz, 
	 $\gamma_2=40$kHz, 
	 $\Omega_\text{c}=1\gamma_4$, $\Omega_\text{s}=0.3\gamma_4$,
	 $\delta_\text{s}=0.5 \Omega_\text{c}$
	 and $\delta_\text{c}=0$.}
 \label{fig:signalplots}
  \end{figure}
The corresponding absorption and dispersion curves for the signal field are plotted in Figs.~\ref{fig:signalplots}.

Similar to the probe-field cases shown in Figs.~\ref{fig:probeplots} and~\ref{fig:Rprobeplots}
we observe two EIT windows in the signal-field absorption plot and gain in the second window.
From Figs.~\ref{fig:probeplots} and~\ref{fig:signalplots} we see that 
the first probe and signal EIT windows are both centered at $\delta_\text{c}$.
The second EIT windows are centered at $\delta_\text{p}=\delta_\text{s}$,
which differs from~$\delta_\text{c}$.
Gain is present in each of the linear susceptibilities for signal and probe second EIT windows.

Simultaneous slowing of beams and matching their group velocities
is advantageous for enhancing inter-beam interactions, such as for cross-phase modulation.
Here we have two transparency windows for each of the signal and probe beams.
Now we seek to match
group velocities for each of the signal and probe fields such that the velocities match for both for the first EIT window and also for the second EIT window.
In other words, we wish to have DEIT for the signal and probe for the first EIT windows,
which is the usual case of DEIT,
and also to have DEIT for the second EIT windows of each of the signal and probe,
where gain occurs.
Through this DDEIT phenomenon,
one could send bichromatic signal and probe fields through the medium
with the lower-frequency chromatic component of the signal and probe fields travelling with
one simultaneously matched group velocity
and the upper-frequency chromatic component also travelling
through the medium at a different but simultaneously matched group velocity.

For group-velocity matching in $^{87}$Rb with the same parameters as before,
we choose field strengths
$\Omega_\text{s}=0.300\gamma_4$,
$\Omega_\text{p}=0.245\gamma_4$ and
$\Omega_\text{c}=1.000\gamma_4$.
The resultant group velocities are 
nearly identical at $30.10\text{ms}^{-1}$ and $30.13\text{ms}^{-1}$
for the probe and signal fields, respectively, in the first EIT window,
and $0.91\text{ms}^{-1}$ for both probe and signal field in the second EIT window.
The two EIT windows for each of the signal and probe are separated by $\Omega_\text{c}/2=9.00$MHz
with a first-window FWHM of 10.50MHz and a second-window width of 1.50MHz.

The second window is quite narrow but experimentally resolvable.
This second EIT window for the probe has been observed for the $^{87}$Rb D1 line
although the width and other features of this window were not investigated~\cite{Li2007}.
Thus, simultaneous matching of signal and probe group velocities in each of the two EIT windows should be possible
with the reasonable experimental parameters.

Thus far we assumed natural linewidths but now consider robustness
subject to driving-field linewidth broadening and temperature-dependent Doppler broadening.
Laser linewidth broadening dephases atomic transitions
but does not modify atomic populations~\cite{sadaf1994},
hence is accounted for by a dephasing-rate replacement~\cite{LBWX97,Li2007}
$\gamma_\phi\to\gamma_\phi+\Delta$
in Eq.~(\ref{eq:masterequation})
with~$\Delta$ the full width at half maximum of the laser line.
Specifically we modify the homogeneous dephasing rates by the laser broadened dephasing rates
for each of the probe, coupling and signal fields and assume independence of all driving-field sources.

Doppler broadening is accounted for by averaging the complex susceptibility
over a Maxwell distribution of velocities
such that the root-mean-square atomic velocity is $\sqrt{2kT/m}$ for $m$ the mass of Rubidium, $k$ Boltzmann's constant and $T$ the temperature in Kelvin~\cite{ Julio1995}.
For the D1 line of~$^{87}$Rb,
two-photon transitions are completely Doppler-free because
the three driving fields drive approximately equal transition frequencies:
$\omega_p\approx\omega_c\approx\omega_s $.
Therefore each~$\delta_\text{xy}$ in Eqs.~(\ref{eq:rho14}) and~(\ref{eq:rho34}) does not change
under Doppler broadening.

\begin{figure}
 \centering
\setbox1=\hbox{\includegraphics[height=5cm]{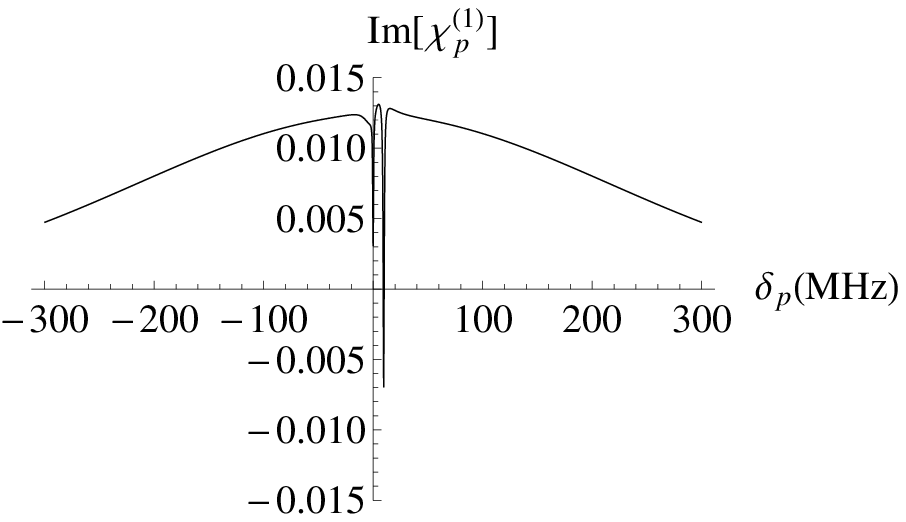}}
  \includegraphics[height=5cm]{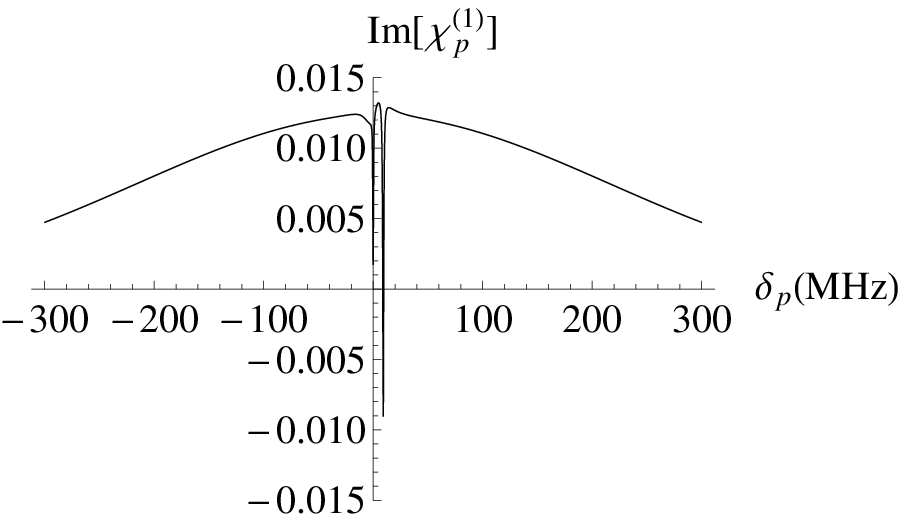}\llap{\raisebox{3.4cm}{\includegraphics[height=2.2cm]{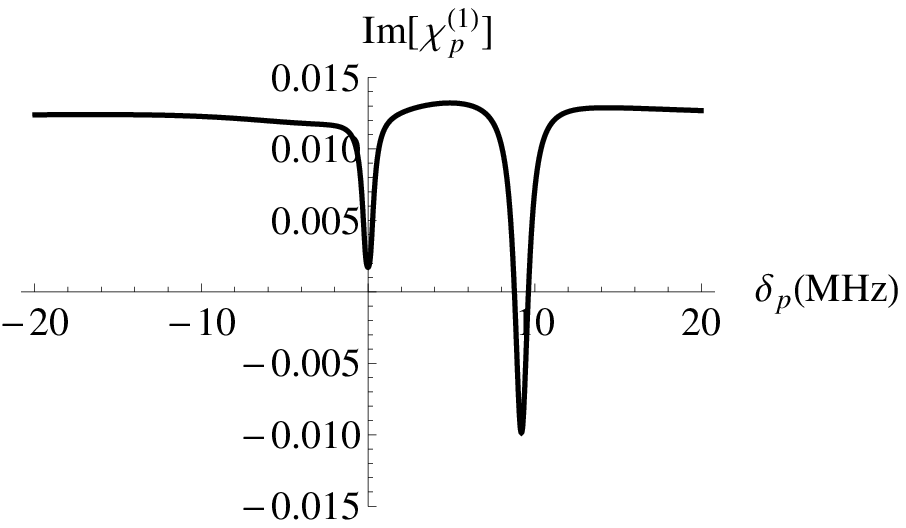}}}
  \caption{Im[$\chi^{(1)}_{p}$] as a function of probe detuning $\delta_\text{p}$
	with $\gamma_4=18$MHz, $\gamma_2=40$kHz,
	 $\Omega_\text{c}=1\gamma_4$, $\Omega_\text{s}=0.3\gamma_4$, $\delta_\text{s}=0.5 \Omega_\text{c}$ 
	 and $\delta_\text{c}=0$.
	 Laser linewidths are $\Delta_p=\Delta_c=\Delta_s=0.1$MHz and $T=$300K. 
	 Inset:\ magnification showing EIT and gain.}
\label{fig:DOPdop}
\end{figure}

We compute the imaginary part of susceptibility using the same parameters as in Fig.~\ref{fig:probeplots}
but for room temperature, i.e., 300K, and show the result in Fig.~\ref{fig:DOPdop}.
Comparing these figures shows a reduction of EIT window width commensurate with past observations~\cite{YZ2002}.
Furthermore we observe a reduction in transparency and gain due to both the driving-field linewidth effect~\cite{LBWX97} and Doppler broadening~\cite{Vemuri96}.
Despite Doppler broadening, both windows are still evident,
and the second window is evidently more robust than the first with respect to Doppler broadening.
The narrowness of the transparency window imposed by Doppler broadening will produce an ultra-slow
group velocity~\cite{Kash1999}.

In summary we have shown that a tripod ($\pitchfork$) electronic energy structure in a four-level atom
can yield rich hitherto unstudied phenomena, in particular double-double electromagnetically induced
transparency with gain.
We have used a  semi-classical-dressed picture to connect the $\pitchfork$ electronic structure to a 
double-$\Lambda$ electronic structure to explain how each of a signal and probe field
experience double-EIT windows.
In the well studied case of DEIT,
a signal and probe would each have an EIT window such that both fields can be slowed at the
same time and also could interact via cross-phase modulation.
In our case DDEIT exhibits DEIT for both the first EIT windows of the signal and probe and
also for the second windows.

Our DDEIT scheme should be experimentally feasible and we have employed realistic
parameters for $^{87}$Rb including driving-laser linewidths and temperature.
Previous observations of one aspect of our scheme,
namely the second window for the EIT probe~\cite{Li2007,MCL08},
reinforce that this scheme is within experimental reach.
DEIT for the second EIT windows for the signal and probe fields do not just replicate the nature of the
first because the second windows also show coherent gain,
and our scheme could be especially interesting for controlling bichromatic signal and probe fields.

\acknowledgments
We acknowledge valuable discussions with S.\ Rebic and P.\ Anisimov
and support from AITF, CIFAR, PIMS, NSERC and the China Thousand Talents Program.
\bibliography{ddeit}

\begin{thebibliography}{20}
\expandafter\ifx\csname natexlab\endcsname\relax\def\natexlab#1{#1}\fi
\expandafter\ifx\csname bibnamefont\endcsname\relax
  \def\bibnamefont#1{#1}\fi
\expandafter\ifx\csname bibfnamefont\endcsname\relax
  \def\bibfnamefont#1{#1}\fi
\expandafter\ifx\csname citenamefont\endcsname\relax
  \def\citenamefont#1{#1}\fi
\expandafter\ifx\csname url\endcsname\relax
  \def\url#1{\texttt{#1}}\fi
\expandafter\ifx\csname urlprefix\endcsname\relax\def\urlprefix{URL }\fi
\providecommand{\bibinfo}[2]{#2}
\providecommand{\eprint}[2][]{\url{#2}}

\bibitem[{\citenamefont{Harris}(1997)}]{Harris1997}
\bibinfo{author}{\bibfnamefont{S.~E.} \bibnamefont{Harris}},
  \bibinfo{journal}{Phys. Today} \textbf{\bibinfo{volume}{50}},
  \bibinfo{pages}{36} (\bibinfo{year}{1997}).

\bibitem[{\citenamefont{Lvovsky et~al.}(2009)\citenamefont{Lvovsky, Sanders,
  and Tittel}}]{Lvovsky2009}
\bibinfo{author}{\bibfnamefont{A.~I.} \bibnamefont{Lvovsky}},
  \bibinfo{author}{\bibfnamefont{B.~C.} \bibnamefont{Sanders}},
  \bibnamefont{and} \bibinfo{author}{\bibfnamefont{W.}~\bibnamefont{Tittel}},
  \bibinfo{journal}{Nature Photonics} \textbf{\bibinfo{volume}{3}},
  \bibinfo{pages}{706 } (\bibinfo{year}{2009}).

\bibitem[{\citenamefont{Rebic et~al.}(2004)\citenamefont{Rebic, Vitali,
  Ottaviani, Tombesi, Artoni, Cataliotti, and Corbalan}}]{Rebic2004}
\bibinfo{author}{\bibfnamefont{S.}~\bibnamefont{Rebic}},
  \bibinfo{author}{\bibfnamefont{D.}~\bibnamefont{Vitali}},
  \bibinfo{author}{\bibfnamefont{C.}~\bibnamefont{Ottaviani}},
  \bibinfo{author}{\bibfnamefont{P.}~\bibnamefont{Tombesi}},
  \bibinfo{author}{\bibfnamefont{M.}~\bibnamefont{Artoni}},
  \bibinfo{author}{\bibfnamefont{F.}~\bibnamefont{Cataliotti}},
  \bibnamefont{and} \bibinfo{author}{\bibfnamefont{R.}~\bibnamefont{Corbalan}},
  \bibinfo{journal}{Phys. Rev. A} \textbf{\bibinfo{volume}{70}},
  \bibinfo{pages}{032 317} (\bibinfo{year}{2004}).

\bibitem[{\citenamefont{Joshi and Xiao}(2005)}]{Joshi2005}
\bibinfo{author}{\bibfnamefont{A.}~\bibnamefont{Joshi}} \bibnamefont{and}
  \bibinfo{author}{\bibfnamefont{M.}~\bibnamefont{Xiao}},
  \bibinfo{journal}{Phys. Rev. A} \textbf{\bibinfo{volume}{72}},
  \bibinfo{pages}{062319} (\bibinfo{year}{2005}),
  \urlprefix\url{http://link.aps.org/doi/10.1103/PhysRevA.72.062319}.

\bibitem[{\citenamefont{Wang et~al.}(2006)\citenamefont{Wang, Marzlin, and
  Sanders}}]{Wang2006}
\bibinfo{author}{\bibfnamefont{Z.-B.} \bibnamefont{Wang}},
  \bibinfo{author}{\bibfnamefont{K.-P.} \bibnamefont{Marzlin}},
  \bibnamefont{and} \bibinfo{author}{\bibfnamefont{B.~C.}
  \bibnamefont{Sanders}}, \bibinfo{journal}{Phys. Rev. Lett.}
  \textbf{\bibinfo{volume}{97}}, \bibinfo{pages}{063901}
  (\bibinfo{year}{2006}),
  \urlprefix\url{http://link.aps.org/doi/10.1103/PhysRevLett.97.063901}.

\bibitem[{\citenamefont{Ottaviani et~al.}(2006)\citenamefont{Ottaviani,
  Rebi\ifmmode~\acute{c}\else \'{c}\fi{}, Vitali, and Tombesi}}]{Ottaviani2006}
\bibinfo{author}{\bibfnamefont{C.}~\bibnamefont{Ottaviani}},
  \bibinfo{author}{\bibfnamefont{S.}~\bibnamefont{Rebi\ifmmode~\acute{c}\else
  \'{c}\fi{}}}, \bibinfo{author}{\bibfnamefont{D.}~\bibnamefont{Vitali}},
  \bibnamefont{and} \bibinfo{author}{\bibfnamefont{P.}~\bibnamefont{Tombesi}},
  \bibinfo{journal}{Phys. Rev. A} \textbf{\bibinfo{volume}{73}},
  \bibinfo{pages}{010301} (\bibinfo{year}{2006}),
  \urlprefix\url{http://link.aps.org/doi/10.1103/PhysRevA.73.010301}.

\bibitem[{\citenamefont{MacRae et~al.}(2008)\citenamefont{MacRae, Campbell, and
  Lvovsky}}]{MCL08}
\bibinfo{author}{\bibfnamefont{A.}~\bibnamefont{MacRae}},
  \bibinfo{author}{\bibfnamefont{G.}~\bibnamefont{Campbell}}, \bibnamefont{and}
  \bibinfo{author}{\bibfnamefont{A.~I.} \bibnamefont{Lvovsky}},
  \bibinfo{journal}{Opt. Lett} \textbf{\bibinfo{volume}{33}},
  \bibinfo{pages}{2659} (\bibinfo{year}{2008}).

\bibitem[{\citenamefont{Berman}(1996)}]{Ber96}
\bibinfo{author}{\bibfnamefont{P.~R.} \bibnamefont{Berman}},
  \bibinfo{journal}{Phys. Rev. A} \textbf{\bibinfo{volume}{53}},
  \bibinfo{pages}{2627} (\bibinfo{year}{1996}),
  \urlprefix\url{http://link.aps.org/doi/10.1103/PhysRevA.53.2627}.

\bibitem[{\citenamefont{Fleischhauer et~al.}(2005)\citenamefont{Fleischhauer,
  Imamoglu, and Marangos}}]{Fleischhauer2005}
\bibinfo{author}{\bibfnamefont{M.}~\bibnamefont{Fleischhauer}},
  \bibinfo{author}{\bibfnamefont{A.}~\bibnamefont{Imamoglu}}, \bibnamefont{and}
  \bibinfo{author}{\bibfnamefont{J.~P.} \bibnamefont{Marangos}},
  \bibinfo{journal}{Rev. Mod. Phys.} \textbf{\bibinfo{volume}{77}},
  \bibinfo{pages}{633} (\bibinfo{year}{2005}),
  \urlprefix\url{http://link.aps.org/doi/10.1103/RevModPhys.77.633}.

\bibitem[{\citenamefont{Vewinger et~al.}(2007)\citenamefont{Vewinger, Appel,
  Figueroa, and Lvovsky}}]{VAFL07}
\bibinfo{author}{\bibfnamefont{F.}~\bibnamefont{Vewinger}},
  \bibinfo{author}{\bibfnamefont{J.}~\bibnamefont{Appel}},
  \bibinfo{author}{\bibfnamefont{E.}~\bibnamefont{Figueroa}}, \bibnamefont{and}
  \bibinfo{author}{\bibfnamefont{A.~I.} \bibnamefont{Lvovsky}},
  \bibinfo{journal}{Opt. Lett.} \textbf{\bibinfo{volume}{32}},
  \bibinfo{pages}{2771} (\bibinfo{year}{2007}).

\bibitem[{\citenamefont{Paspalakis and Knight}(2002)}]{Paspalakis2002}
\bibinfo{author}{\bibfnamefont{E.}~\bibnamefont{Paspalakis}} \bibnamefont{and}
  \bibinfo{author}{\bibfnamefont{P.~L.} \bibnamefont{Knight}},
  \bibinfo{journal}{Phys. Rev. A} \textbf{\bibinfo{volume}{66}},
  \bibinfo{pages}{015802} (\bibinfo{year}{2002}).

\bibitem[{\citenamefont{Li et~al.}(2007)\citenamefont{Li, Yang, Cao, Zhang, and
  Xie}}]{Li2007}
\bibinfo{author}{\bibfnamefont{S.}~\bibnamefont{Li}},
  \bibinfo{author}{\bibfnamefont{X.}~\bibnamefont{Yang}},
  \bibinfo{author}{\bibfnamefont{X.}~\bibnamefont{Cao}},
  \bibinfo{author}{\bibfnamefont{C.}~\bibnamefont{Zhang}}, \bibnamefont{and}
  \bibinfo{author}{\bibfnamefont{H.}~\bibnamefont{Xie},
  \bibfnamefont{C.and~Wang}}, \bibinfo{journal}{Phys. B: At. Mol. Opt. Phys.}
  \textbf{\bibinfo{volume}{40}}, \bibinfo{pages}{3211} (\bibinfo{year}{2007}).

\bibitem[{\citenamefont{Kocharovskaya}(1992)}]{Koc92}
\bibinfo{author}{\bibfnamefont{O.}~\bibnamefont{Kocharovskaya}},
  \bibinfo{journal}{Phys. Rep.} \textbf{\bibinfo{volume}{219}},
  \bibinfo{pages}{175} (\bibinfo{year}{1992}).

\bibitem[{\citenamefont{Cohen and Berman}(1996)}]{CB96}
\bibinfo{author}{\bibfnamefont{J.~L.} \bibnamefont{Cohen}} \bibnamefont{and}
  \bibinfo{author}{\bibfnamefont{P.~R.} \bibnamefont{Berman}},
  \bibinfo{journal}{Phys. Rev. A} \textbf{\bibinfo{volume}{55}},
  \bibinfo{pages}{3900} (\bibinfo{year}{1996}).

\bibitem[{\citenamefont{Sultana and Zubairy}(1994)}]{sadaf1994}
\bibinfo{author}{\bibfnamefont{S.}~\bibnamefont{Sultana}} \bibnamefont{and}
  \bibinfo{author}{\bibfnamefont{M.~S.} \bibnamefont{Zubairy}},
  \bibinfo{journal}{Phys. Rev. A} \textbf{\bibinfo{volume}{49}},
  \bibinfo{pages}{438} (\bibinfo{year}{1994}),
  \urlprefix\url{http://link.aps.org/doi/10.1103/PhysRevA.49.438}.

\bibitem[{\citenamefont{L\"u et~al.}(1997)\citenamefont{L\"u, Burkett, and
  Xiao}}]{LBWX97}
\bibinfo{author}{\bibfnamefont{B.}~\bibnamefont{L\"u}},
  \bibinfo{author}{\bibfnamefont{W.~H.} \bibnamefont{Burkett}},
  \bibnamefont{and} \bibinfo{author}{\bibfnamefont{M.}~\bibnamefont{Xiao}},
  \bibinfo{journal}{Phys. Rev. A} \textbf{\bibinfo{volume}{56}},
  \bibinfo{pages}{976} (\bibinfo{year}{1997}),
  \urlprefix\url{http://link.aps.org/doi/10.1103/PhysRevA.56.976}.

\bibitem[{\citenamefont{Gea-Banacloche
  et~al.}(1995)\citenamefont{Gea-Banacloche, Li, Jin, and Xiao}}]{Julio1995}
\bibinfo{author}{\bibfnamefont{J.}~\bibnamefont{Gea-Banacloche}},
  \bibinfo{author}{\bibfnamefont{Y.-q.} \bibnamefont{Li}},
  \bibinfo{author}{\bibfnamefont{S.-z.} \bibnamefont{Jin}}, \bibnamefont{and}
  \bibinfo{author}{\bibfnamefont{M.}~\bibnamefont{Xiao}},
  \bibinfo{journal}{Phys. Rev. A} \textbf{\bibinfo{volume}{51}},
  \bibinfo{pages}{576} (\bibinfo{year}{1995}),
  \urlprefix\url{http://link.aps.org/doi/10.1103/PhysRevA.51.576}.

\bibitem[{\citenamefont{Ye and Zibrov}(2002)}]{YZ2002}
\bibinfo{author}{\bibfnamefont{C.~Y.} \bibnamefont{Ye}} \bibnamefont{and}
  \bibinfo{author}{\bibfnamefont{A.~S.} \bibnamefont{Zibrov}},
  \bibinfo{journal}{Phys. Rev. A} \textbf{\bibinfo{volume}{65}},
  \bibinfo{pages}{023806} (\bibinfo{year}{2002}),
  \urlprefix\url{http://link.aps.org/doi/10.1103/PhysRevA.65.023806}.

\bibitem[{\citenamefont{Vemuri and Agarwal}(1996)}]{Vemuri96}
\bibinfo{author}{\bibfnamefont{G.}~\bibnamefont{Vemuri}} \bibnamefont{and}
  \bibinfo{author}{\bibfnamefont{G.~S.} \bibnamefont{Agarwal}},
  \bibinfo{journal}{Phys. Rev. A} \textbf{\bibinfo{volume}{53}},
  \bibinfo{pages}{1060} (\bibinfo{year}{1996}),
  \urlprefix\url{http://link.aps.org/doi/10.1103/PhysRevA.53.1060}.

\bibitem[{\citenamefont{Kash et~al.}(1999)\citenamefont{Kash, Sautenkov,
  Zibrov, Hollberg, Welch, Lukin, Rostovtsev, Fry, and Scully}}]{Kash1999}
\bibinfo{author}{\bibfnamefont{M.~M.} \bibnamefont{Kash}},
  \bibinfo{author}{\bibfnamefont{V.~A.} \bibnamefont{Sautenkov}},
  \bibinfo{author}{\bibfnamefont{A.~S.} \bibnamefont{Zibrov}},
  \bibinfo{author}{\bibfnamefont{L.}~\bibnamefont{Hollberg}},
  \bibinfo{author}{\bibfnamefont{G.~R.} \bibnamefont{Welch}},
  \bibinfo{author}{\bibfnamefont{M.~D.} \bibnamefont{Lukin}},
  \bibinfo{author}{\bibfnamefont{Y.}~\bibnamefont{Rostovtsev}},
  \bibinfo{author}{\bibfnamefont{E.~S.} \bibnamefont{Fry}}, \bibnamefont{and}
  \bibinfo{author}{\bibfnamefont{M.~O.} \bibnamefont{Scully}},
  \bibinfo{journal}{Phys. Rev. Lett.} \textbf{\bibinfo{volume}{82}},
  \bibinfo{pages}{5229} (\bibinfo{year}{1999}),
  \urlprefix\url{http://link.aps.org/doi/10.1103/PhysRevLett.82.5229}.

\end{thebibliography}
\end{document}